# Gas transport mechanisms through molecularly thin membranes

V. Stroganov[1], D. Hüger[1], C. Neumann[1], T. Noethel[1], M. Steinert[2], U. Hübner[3], A. Turchanin[1*]

[1]*Institute of Physical Chemistry, Friedrich Schiller University Jena, 07743 Jena, Germany*
[2]*Institute of Applied Physics, Friedrich Schiller University Jena, 07743 Jena, Germany*
[3]*Leibniz Institute of Photonic Technology (IPHT), 07745 Jena, Germany*

Corresponding author: andrey.turchanin@uni-jena.de

**Atomically thin molecular carbon nanomembranes (CNMs) with intrinsic sub-nanometer porosity are considered as promising candidates for next generation filtration and gas separation applications due to their extremely low thickness, energy efficiency and selectivity[1, 2, 3, 4, 5]. The porous structure of CNMs gives them advantage over other 2D materials such as graphene and transition metal dichalcogenides where defects and pores need to be introduced after synthesis[6, 7, 8]. Previous study of gas permeation through 4,4'-terphenylthiol (TPT) CNM showed only helium and water vapour permeation above the limit of detection[3]. The permeation of water vapour was nonlinear against its pressure and 1000 stronger than permeation of helium despite their similar kinetic diameters. This anomalous behavior was explained by cooperative movement and clustering of water molecules due to hydrogen bonding[9]. However, the exact mechanisms remained unknown. Here, we demonstrate that the character of permeation is defined by adsorption of gas species. We performed gas permeation measurements through TPT CNM at different temperatures and found that all measured gases experienced an activation energy barrier which correlated with their kinetic diameters. Furthermore, we identified that entropy loss during adsorption and permeation is the fundamental reason of strong nonlinear permeation of water. Our results also demonstrated that adsorption plays a major role in permeation of all gases, not just water.**

Modern world constantly needs new solutions for filtration and gas separation applications. CNMs have thickness of around 1 nm which lowers the required pressure difference and energy to maintain adequate flux across the membrane during filtration processes[1]. However, the precise understanding about the permeation mechanism through these materials remains unknown[9]. The motivation behind this work was to explain strong nonlinear permeation of water[3], which then evolved into a full-scale investigation of gas permeation mechanism through CNMs. Initial literature overview revealed that the character of permeation of water through TPT CNM[3, 9] strongly resembled the subtype of Brunauer-Emmet-Teller (BET) adsorption isotherm[10, 11] characteristic for adsorbates with stronger affinity to themselves than to a substrate. This behavior is expected to be the case for water adsorption on TPT CNM because of its hydrophobic nature. Additionally, similar non-linear permeation of water and other vapors was routinely observed for traditional polymer membranes[12, 13] and is described by the solution-diffusion model[14]. This model states that gases first dissolve in a membrane and then diffuse through it following the difference in chemical potentials. However, the term "solution" is hardly applicable in case of CNMs because of their extremely low nanometer thickness. Instead, we propose an analogous mechanism called adsorption-diffusion which was described earlier for nanoporous graphene[7, 8]. According to it, the diffusion occurs from the adsorbed layer of gas molecules (Figure 1a). Thus, permeation of water vapour and other gases is defined by the character of their adsorption. Additionally, permeating species should experience activation energy barrier during diffusion which should correlate with their respective kinetic diameters. In this work we investigated gas permeation through TPT CNM at different temperatures to evaluate the applicability of the adsorption-diffusion model and explain differences in permeation of water and other gases.

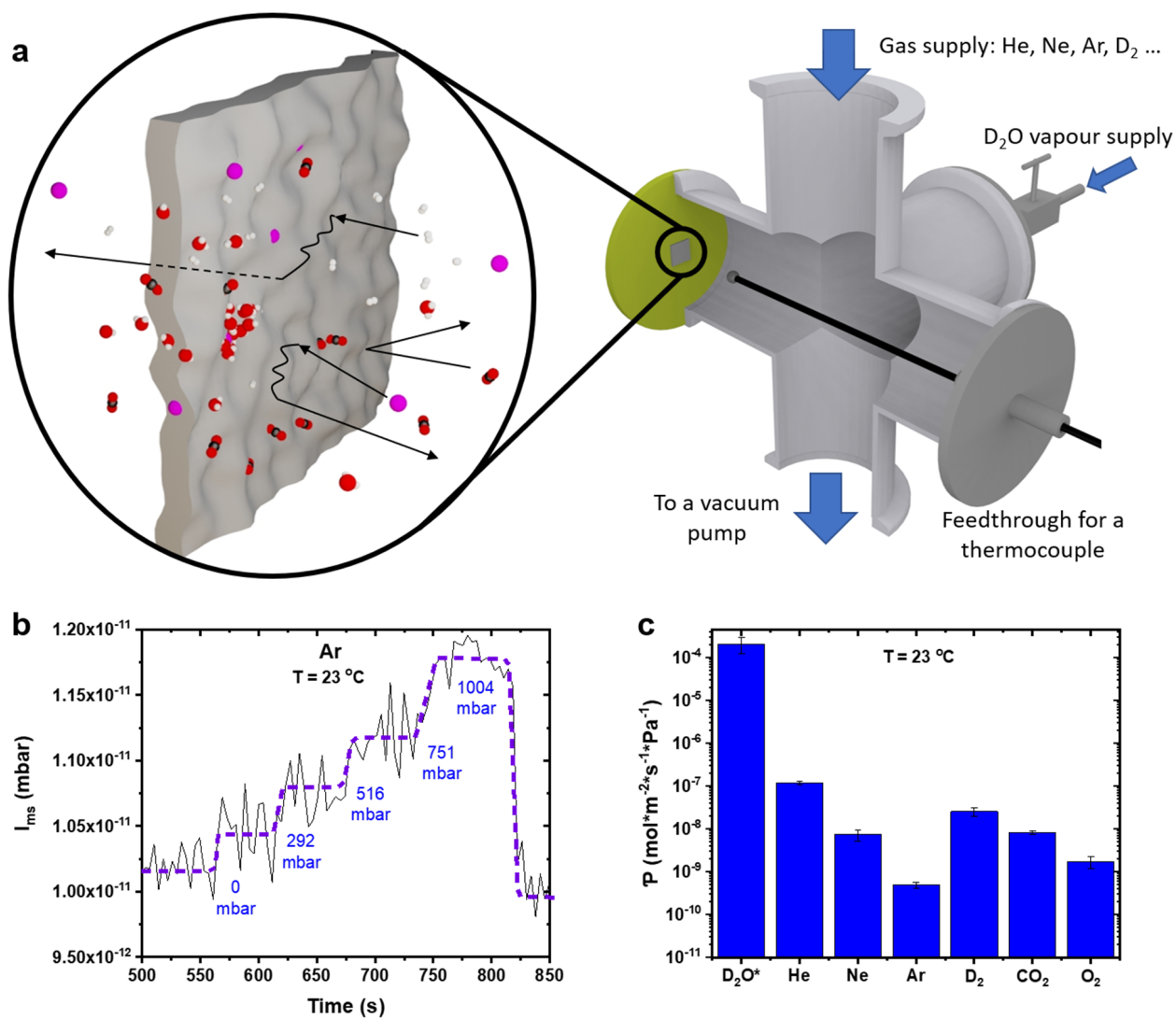

***Figure 1. The model of the gas permeation and the summary of the results.***
***a** The left part shows the model of a CNM and gas permeation mechanism. Black arrows and lines show potential paths molecules can take upon interaction with the CNM: adsorption-diffusion, adsorption-desorption or deflection. The right part shows the general scheme of the permeation experiment. Sample with TPT CNM is fixed in a setup separating gas chamber from the vacuum chamber with mass spectrometer. The gas chamber and the sample fixture are heated by an external heat source. See more details in the supplementary Figure S1. **b** The raw data of permeation of argon measured by the mass spectrometer at room temperature. The blue dashed line is to guide the eye. Pressure values next to each "step" represent the applied argon pressure in the gas chamber during measurement. **c** Summary of gas and vapour permeances at room temperature for TPT CNM. The average value is taken across 3 samples and error bars denote the standard deviation. Noble and multiatomic gases are grouped to emphasize the correlation between their permeances and molecular kinetic diameters. *$D_2O$ is an exception from this trend: its permeance is nonlinear against its vapour pressure; the value shown here is taken at 24 mbar which is close to saturated vapour pressure.*

The measurements of gas permeation were performed using a mass spectrometer separated by the CNM sample from the gas chamber. In this setup, the gas transport from gas to vacuum chambers was only possible through the CNM. The detailed description of the experimental apparatus can be found in the Supplementary Information.

At first, we performed measurements of gas permeation at room temperature (23 °C) and were able to detect permeation of argon through TPT CNM (Figure 1b) for the first time. The figure demonstrates that the permeation was extremely weak and signal-to-noise ratio was low. However, the change of the average signal with increased supply pressure was linear which indicated that we measured the real permeation instead of a random noise. This detection became possible thanks to the measures to reduce the total pressure in the vacuum chamber down to ~6×10$^{-10}$ mbar. After argon, we tested permeation of helium, neon, deuterium, carbon dioxide, oxygen and deuterium oxide at room temperature. The results are shown in Figure 1c. One can see that permeance of deuterium oxide was ~1000 times stronger than permeance of helium and other gases. This result is consistent with the previous study[3]. It is also notable that permeance of noble gases decreased from helium to argon. The same tendency is observed for deuterium, carbon dioxide and oxygen. This decrease could be attributed either to lower impingement rate for heavier gases or to the presence of an energy barrier which is defined by kinetic diameter of the respective gases. However, the impingement rate is inversely proportional to the square root of molecular mass, so its effect is too small to explain the differences in permeances.

We proceeded further with measurements of gas permeation at higher temperatures. Figures 2a,b show examples of how signals of helium and deuterium respectively changed with increased temperature. One can see that the signals of mass spectrometer were higher for each subsequent temperature step. The slopes of the lines were proportional to the permeances of the respective gases at given temperatures. The calculated permeance of helium increased from 1.36×10$^{-7}$ mol*m$^{-2}$s$^{-1}$Pa$^{-1}$ at room temperature to 1.76×10$^{-6}$ mol*m$^{-2}$s$^{-1}$Pa$^{-1}$ at 119 °C. The character of this increase can provide information about the transport mechanism. One possibility is that Knudsen diffusion is the dominant transport mechanism for weakly adsorbing gases which was found to be the case for ultrathin polymer membranes[15]. Knudsen diffusion model is applicable for membranes with effective pore diameters bigger than gas molecules under the free molecular flow regime[16]. However, our TPT CNM samples had no significant pores visible in scanning electron microscope (Figure S3b) and earlier studies indicated that only subnanometer pores are potentially present[3, 17]. Additionally, Knudsen diffusion coefficient is proportional to a square root of temperature: $D \sim \sqrt{T}$ [18]. This dependency is weak and cannot explain the increase in the permeance of helium by 1 order of magnitude in the given temperature range. These facts lead us to the conclusion that Knudsen diffusion is unlikely to be the dominant transport mechanism through TPT CNM for weakly adsorbing gases.

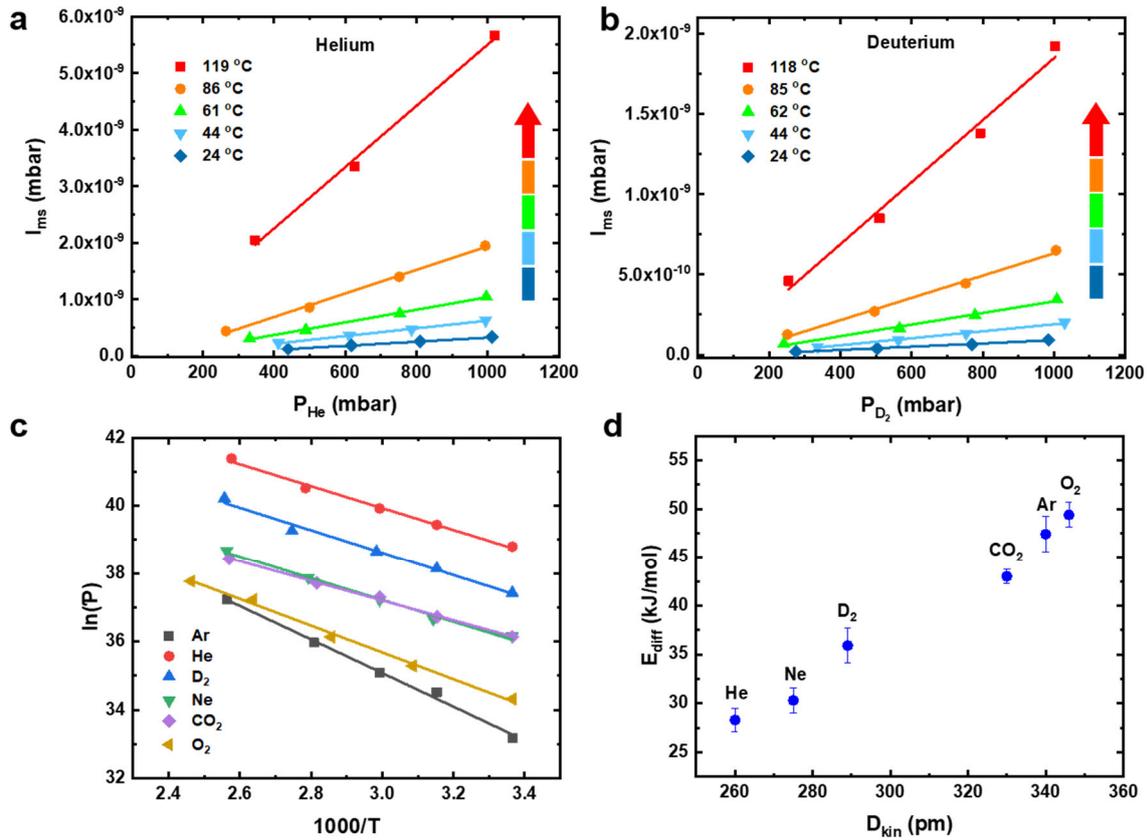

*Figure 2. Temperature dependent permeation of different gases. a, b Dependencies of signal of the mass spectrometer against helium or deuterium feed pressures at different temperatures. c Arrhenius plot of permeances of different gases. d The plot shows correlation between the activation energy of diffusion and kinetic diameters of different gases. The lines in a, b and c are linear fits.*

Another possible explanation of increasing permeation is the presence of an energy barrier similarly to classical dense polymer membranes[19] where permeance followed Arrhenius equation. To prove that our experimental results indeed followed Arrhenius dependency we plotted permeances of gases in Arrhenius coordinates and obtained linear trends (Figure 2c). The slope of each line equals $-\frac{Ea}{R}$, where $E_a$ is the apparent activation energy of permeation. The calculated values of $E_a$ for each gas are shown in the supplementary Figure S6. To further analyze these results, we modelled the permeation as a classical two-step process: $M_{gas} \leftrightarrow M_{ads} \rightarrow M_{diff}$. The first step represents the equilibrium between the bulk and adsorbed phases of gas species M. The second step shows diffusion of the adsorbed molecules through the CNM. In total, this process would include two energy barriers: $\Delta H_{ads}$ - the enthalpy of adsorption and $E_{diff}$ - the activation energy of diffusion. The total apparent activation energy $E_a$ would be the sum of these parameters: $E_a = E_{diff} + \Delta H_{ads}$. By using the model of an adsorbed 2D gas developed for graphene[8] we could express the gas permeance as:

$$\mathcal{P} = \mathcal{P}_0 \times exp\left(-\frac{\Delta S}{R}\right) \times exp\left(-\frac{E_{diff}+\Delta H_{ads}}{RT}\right) \quad (1)$$

Here, $\mathcal{P}$ is the permeance, $\mathcal{P}_0$ is the exponential pre-factor weakly dependent on temperature, $\Delta S$ is the entropy loss during the permeation process, R is the universal gas constant and $T$ is the absolute temperature. The equation (1) showed us that the measured apparent activation energy $E_a$ should be lower than the activation energy of diffusion $E_{diff}$ by the value $|\Delta H_{ads}|$. This relation would make it possible to calculate $E_{diff}$ if $\Delta H_{ads}$ is known. To this date however, there were no data about enthalpy of adsorption of gases on CNMs. We suggested that the adsorption of gases on similar carbon materials could be used in (1). This data is shown in supplementary Figure S6. Figure 2d shows the final calculated values of $E_{diff}$ plotted against kinetic diameters of the respective gases. One can see a clear correlation – gases with bigger kinetic diameters experienced higher activation barriers. This correlation suggests that the dominant gas transport mechanism is adsorption-diffusion. Furthermore, the potential pores present in TPT CNM are subnanometer in size and tortuous[3, 17]. Due to their size and shape, these pores can be considered as free volume inside a dense CNM similar to free volume space in amorphous polymers[20]. Under these assumptions, ballistic transport of gas molecules through any dense CNM is not possible. The only possible pathway would be *via* adsorption and diffusion making adsorption an important step in the permeation process. The linear character of permeation of weakly adsorbing gases (Figure 2 a, b) can be explained by their linear adsorption isotherms in the given pressure and temperature range. Such adsorption isotherms are characteristic for Henry adsorption[21].

In contrast to other gases, $D_2O$ had nonlinear permeation. As we mentioned at the beginning, the shape of the curve of $D_2O$ permeation at room temperature resembled BET adsorption isotherm (Figure 3a, 23°C). We proposed that this dependence is not a coincidence but rather the permeation curve of $D_2O$ reflects its adsorption isotherm on TPT CNM. It was shown[8] that significant entropy loss of the permeation process can lead to enhanced molecular flows which is also reflected in the equation (1). One can expect that $D_2O$ molecules experience stronger entropy loss than helium atoms during adsorption and diffusion which explains much stronger permeation of water. Additionally, it was shown[3] that the permeation of water at saturated vapor pressure is similar to the permeation of liquid water. These findings indicate that condensation of water vapor happened on the CNM[9] leading to increased permeation. The equation (1) can be used to give the fundamental explanation to this phenomenon. When the partial pressure of $D_2O$ approached saturation, condensation of water lead to even stronger entropy loss than during regular adsorption. This resulted in the exponential increase of the permeation by the factor $\exp(|\Delta S|/R)$. This phenomenon is known as "crowding effect"[8] and is observed for strongly adsorbing species. To further strengthen the argument that entropy loss explains the difference in permeation of water and other gases we estimated the enhancement factor for $D_2O$ against $CO_2$ based on the data of their adsorption on zeolites[22]. The calculation resulted in the factor of $10^5$ which is close to the experimentally observed difference in their permeation presented in this work.

Further measurements at higher temperature showed decreasing permeation of $D_2O$. This behavior can be explained by the negative $E_a$ which in turn suggested that $|E_{diff}|<|\Delta H_{ads}|$. This falls perfectly in line with the nature of small strongly-adsorbing $D_2O$ molecules and literature data. Kinetic diameter of $D_2O$ molecule is 265 pm which is close to 260 pm of He atom[23]. This puts $E_{diff}$ of $D_2O$ around 29 kJ/mol. Literature values of enthalpy of adsorption of water on carbon materials are located between -69 and -45 kJ/mol[24, 25]. These data give negative sum of $E_{diff}$ and $\Delta H_{ads}$ for $D_2O$ which leads to decreased permeation at higher temperatures.

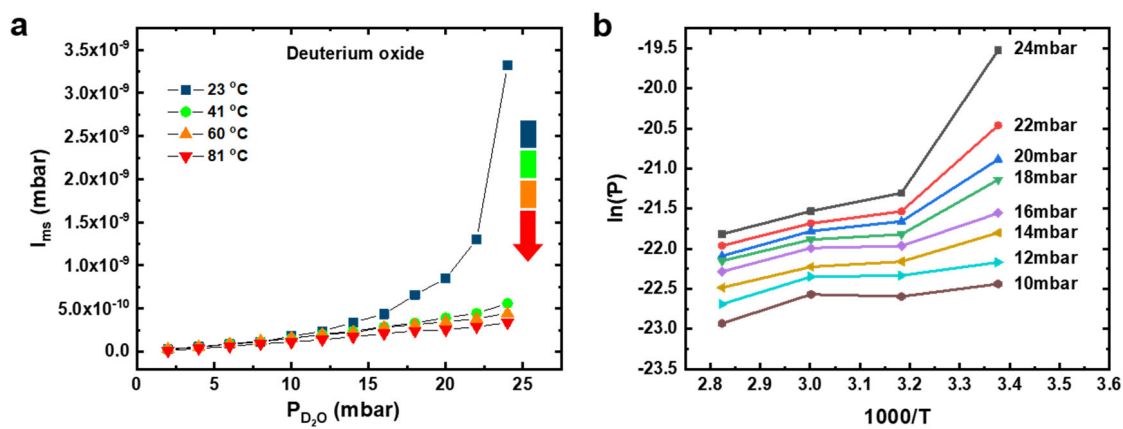

*Figure 3. Temperature-dependent permeation of deuterium oxide vapour. a Influence of feed pressure of $D_2O$ vapour on signal of the mass spectrometer at different temperatures. The permeation of water is nonlinear against vapour pressure. b The Arrhenius plot of permeance of heavy water against temperature at different feed pressures. The lines in a and b are to guide the eye.*

It is known that Arrhenius plots of permeation of substances below their critical temperatures $T < T_c$ are nonlinear in classical solution-diffusion model due to high condensability of the penetrant[12, 13, 26]. In that case both $E_a$ and $\Delta S$ are functions of pressure and temperature $E_a = f(P, T)$, $\Delta S = f(P, T)$. In our work we observed the same effect when we have built Arrhenius plots for permeation of $D_2O$. The permeance of $D_2O$ depended on supply pressure thus it is possible to build a single plot. Instead, we group permeance values measured at the same pressures but different temperatures and built Arrhenius plots for each group (Figure 3b). It is noticeable that there is a significant deviation from linearity in the whole pressure range. The calculated mean $E_a$ values were between -10 and -40 kJ/mol. The assumed value of $E_{diff}$ = 29 kJ/mol for $D_2O$ puts the possible $\Delta H_{ads}$ in the range between -39 and -69 kJ/mol which is within limits for known carbon materials.

Commercial filtration and separation processes are often performed at temperatures where the balance between selectivity, permeability and operation costs is reached. Thus, it is important to consider selectivity of separation of different gas pairs at different temperatures. In Figure 4 we summarized the selectivity data for different industrially important gas pairs. One of the most notable pairs is $D_2O/D_2$ with the selectivity of 2870 at room temperature. This value however is calculated for saturated

D$_2$O vapour pressure of ~24 mbar which cannot be higher at this temperature. Usually, water/hydrogen separation is done at increased temperatures. Our experiments showed that selectivity of their separation at 120°C is around 47 meaning that TPT CNM can be a potential candidate for hydrogen drying application during water electrolysis processes. Other pairs also showed promising separation values at room temperatures while somewhat decreased values at 120 °C.

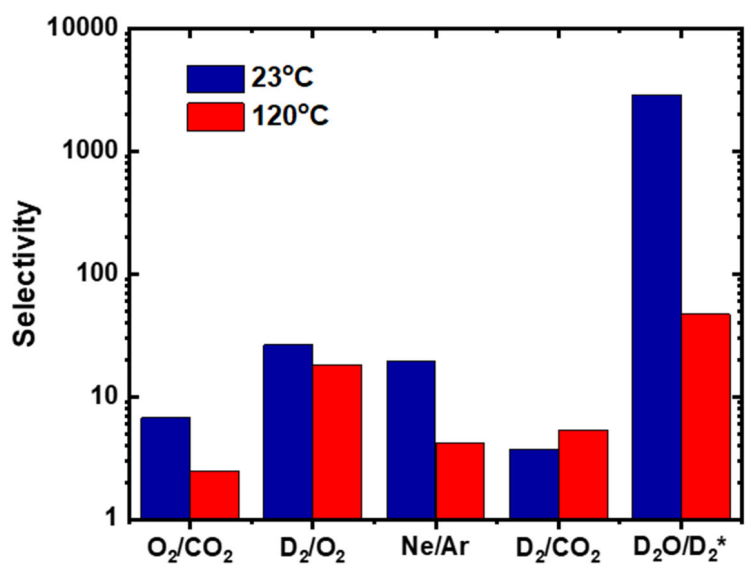

*Figure 4. Separation selectivity.* Selectivities of different gas pairs at room temperature and 120°C. *The selectivity of the D$_2$O/D$_2$ pair was calculated for 24 mbar of D$_2$O vapour pressure.

To conclude, we demonstrated that all gas species experienced energy activation barrier during permeation through TPT CNM. The transport mechanism of all gases involved adsorption step and its parameters played critical role in defining the character of permeation. The significant entropy loss of water molecules was found to be the fundamental explanation for its strong nonlinear permeation.

**Methods**

Preparation of freestanding TPT CNM

TPT self-assembled monolayer (SAM) was synthesized on a 300 nm thick Au layer on a mica substrate (Georg Albert PVD-Coatings). The substrates were cleaned with oxygen plasma and immersed in a dry and degassed solution of TPT in dimethylformamide (DMF) for 24 hours at 70 °C. Resulting TPT SAM was washed with DMF and ethanol and dried under nitrogen flow. It was then crosslinked in high vacuum (< 5*10$^{-8}$ mbar) with an electron gun at an electron energy 50 eV and an average dose of 50 mC/cm$^2$ (FG15/40 Specs). This resulted in a formation of TPT CNM.

Transfer of TPT CNM was performed with a help of a stabilizing ~1 µm thick polymethylmethacrylate (PMMA) layer which was spincoated onto the sample. The Au/TPT/PMMA film was mechanically detached from mica and transferred into a $I_2$/KI bath to etch away Au (5 min). Afterwards the TPT/PMMA film was transferred into a $NaS_2O_3$ bath to remove iodine residues. The film was transferred to water bath twice to remove traces of salts and then transferred to the target substrate with a single hole (Figure S2). The PMMA layer was dissolved by immersing the sample in acetone. Supercritical $CO_2$ was used to extract the sample from the liquid phase without exposing the freestanding CNM to a surface tension[27].

Mass spectrometry

Mass spectrometry measurements were performed on a Hiden Analytical HAL3F-RC mass spectrometer. The scheme of the experimental setup, measurement procedure and evaluation of the results are described in the Supplementary Information.

Optical microscopy

The optical microscopy image was taken with a Zeiss Axio Imager Z1.m microscope equipped with a thermoelectrically cooled 3-megapixel CCD camera (Axiocam 503 color) in bright field operation.

Scanning Electron Microscopy

Scanning electron microscopy (SEM) of the freestanding CNM was performed on a Sigma VP system (Carl Zeiss) at a beam energy of 15 kV using in-lens detector of the system.

**Acknowledgements**

We thank Deutsche Forschungsgemeinschaft DFG for funding *via* the Transregio-Collaborative Research Center TRR234 "Catalight" (Project B7). We acknowledge Stephanie Höppener and Ulrich S. Schubert for providing access to the scanning electron microscope. The SEM facilities of the Jena Center for Soft Matter (JCSM) were established with a grant from the DFG.

**Author contribution**

A.T. conceived and directed the research. V.S. and A.T. deigned the experimental setup and procedures. V.S. and D.H. assembled and implemented the experimental setup. V.S., D.H. and T.N. prepared samples and performed gas permeation measurements. V.S., D.H. and A.T. evaluated and analyzed the experimental data. C.N. conducted SEM. U.H. and M.S. microfabricated $Si_3N_4$ membranes. V.S. and A.T. wrote the manuscript with contributions of all authors.

# Supplementary information

## Gas transport mechanisms through molecularly thin membranes

V. Stroganov[1], D. Hüger[1], C. Neumann[1], T. Noethel[1], M. Steinert[2], U. Huebner[3],

A. Turchanin[1]

[1]*Institute of Physical Chemistry, Friedrich Schiller University Jena, 07743 Jena, Germany*
[2]*Institute of Applied Physics, Friedrich Schiller University Jena, 07743 Jena, Germany*
[3]*Leibniz Institute of Photonic Technology (IPHT), 07745 Jena, Germany*

Corresponding author: andrey.turchanin@uni-jena.de

# 1. Experimental setup

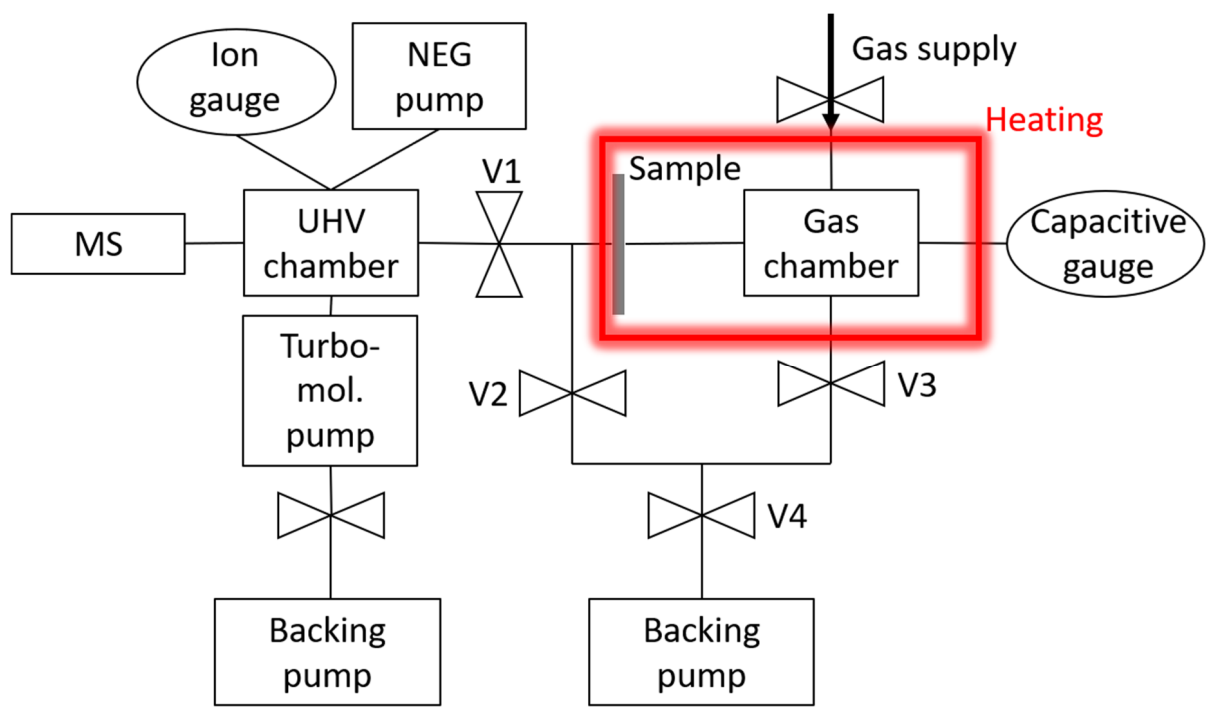

***Figure S1. The experimental setup.*** *The schematic representation of the complete experimental setup. The abbreviations are: MS – mass spectrometer, NEG pump – non-evaporable getter pump, UHV– ultra-high vacuum.*

The experimental setup was assembled around the ultra-high vacuum (UHV) chamber with mass spectrometer. Gas chamber was a stainless steel 6-way KF-40 cross connected to the UHV chamber *via* a KF-CF adapter and CF-35 T-piece. The presence of the KF-CF adapter allowed us to develop 2 alternative sample fixtures which are described in the next section. Installation of a sample was performed by closing valve V1 and disconnecting the gas chamber from the setup. After the installation, the T-piece and the gas chamber contained air which was pumped out by opening valves V2, V3 and V4. The pumping was done for 3 minutes. After that, valve V3 was closed and we waited another 3 minutes to achieve as low pressure as possible in the T-piece. Next, we closed V2 and V4 and then slowly opened V1. The pressure in the UHV chamber usually spiked up to $10^{-5}$ mbar during opening of V1 and then subsequently dropped in the following hours. After V1 was opened we started reactivation of the NEG pump (CapaciTorr Z100, SAES Getters S.p.A.) by attaching a power supply to it which heated the gas adsorbing alloy up to 500 °C for 1 hour. After the sample installation and NEG pump reactivation we left the system overnight to achieve the lowest pressure in the UHV chamber. This was necessary for precise measurements of weakly permeating gases.

The mass spectrometer (MS) used in this work was purchased from Hiden Analytical. The model is HAL3F-RC. It was equipped with a quadrupole mass analyzer with 1-300 amu mass detection range. Electron ionization source with electron energy of 70 eV was used to produce free ions for analysis. The detectors were faraday cup and secondary electron multiplier with sensitivities down to $10^{-10}$ mbar and $10^{-13}$ mbar respectively. The measurements were performed as following. First, the device was setup to measure signal at a specific mass. The signal was recorded until the background was stable. Simultaneously, we opened V3 and V4 to completely empty the gas chamber. When the background was stable enough we closed V3 and slowly opened gas supply valve which allowed for precise manual control of the pressure in the gas chamber. We measured MS signal at different gas supply pressures. This was necessary for further analysis of the results. When the measurement was done we opened V3 to empty the gas chamber. After that, the device was ready to measure permeation of another gas.

Measurements at different temperatures were performed by heating the gas chamber and sample fixtures. This was done using a standard heating belt which was generally purposed for baking of vacuum chambers (Vacom GmbH). The power supply for the heating belt was built in-house and contained an electronic feedback loop to control the surface temperature of the chamber. Detection of the temperature inside the chamber was done using a K-type thermocouple. The feedthrough for K-type thermocouples purchased from Vacom GmbH was used to connect the thermocouple to the outside reading device. The whole heating region was wrapped with aluminum foil to improve heat insulation and achieve more homogeneous temperature distribution. We usually waited 1 hour between setting a new temperature and performing measurements.

The introduction of water vapor was implemented separately from the main gas supply. It was built similar to the way described elsewhere[1]. The container with water was connected to the gas chamber through a needle valve which allowed for precise manual control of the supply pressure. The contents of the container with water were exposed to vacuum before the actual measurement to ensure that there is only pure water vapor above the liquid.

## 2. Sample fixtures

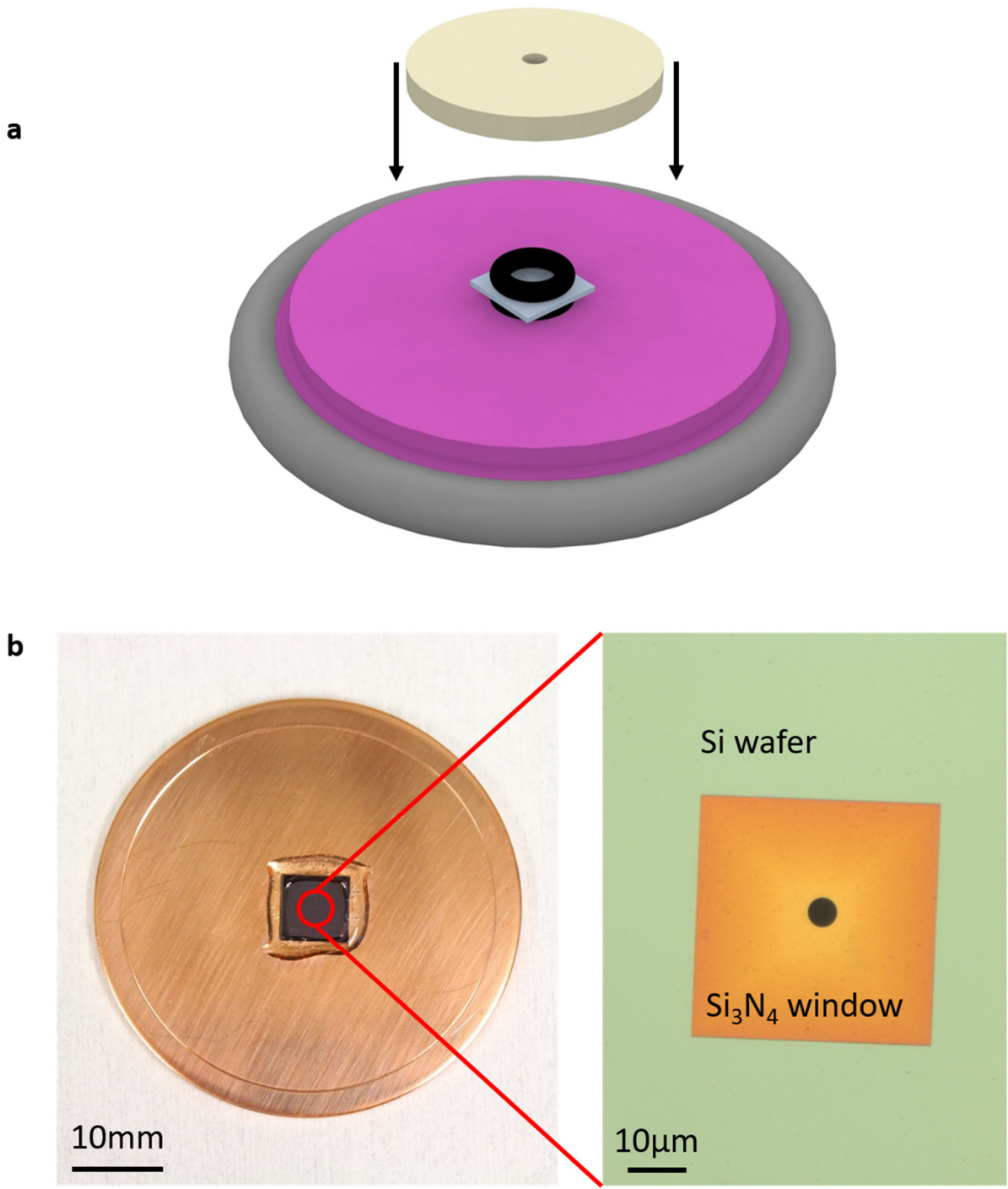

*Figure S2. Sample fixtures. **a** The model of the sample fixture for quick gas leak analysis. KF-40 standard. The sample is shown as the gray square in the middle of the holder. **b** The sample fixture for precise gas permeation measurements. CF-35 standard.*

The target substrate for CNM transfer was a silicon wafer with a 100 nm thick silicon nitride layer. The portion of the silicon wafer was etched to expose the freestanding $Si_3N_4$ membrane. A single hole with 5μm diameter was drilled in it with focused ion beam.

The transfer procedure used in this study usually had a ~50% chance of yielding intact freestanding CNM. To check the integrity of the CNM we designed and built a nonpermanent fixture for KF connection (Figure S2a). Samples were placed between two small vacuum-compatible viton rings and pressed to the body of the holder by the small upper metal circle. This holder allowed for fast mounting and demounting without expendable parts or materials, but it was leaky to helium and couldn't be used for precise measurements. CNM's quality was checked by installing the sample and applying argon pressure. Successful samples which didn't show too strong MS signal for argon were then permanently glued to copper discs with a two-component epoxy resin (Figure S2b) and installed into the system in the CF connection.

### 3. Integrity of CNMs

Preliminary integrity tests were performed by measuring argon leak through the samples as described in the section 2. This method allowed to exclude completely failed samples where freestanding TPT CNM didn't form at all or the aperture was only partially covered. After that, the quality of CNMs was checked by comparing gas permeances between different samples. We measured 3 samples in total and obtained similar values (Figure S3a). This indicated that there were no significant defects in CNMs. To further strengthen this argument, we imaged one sample with scanning electron microscope (Figure S3b) and didn't observe any ruptures. All these measures allowed us to assume that all 3 samples used in this study were of good quality and the gas permeation results could be meaningfully analyzed.

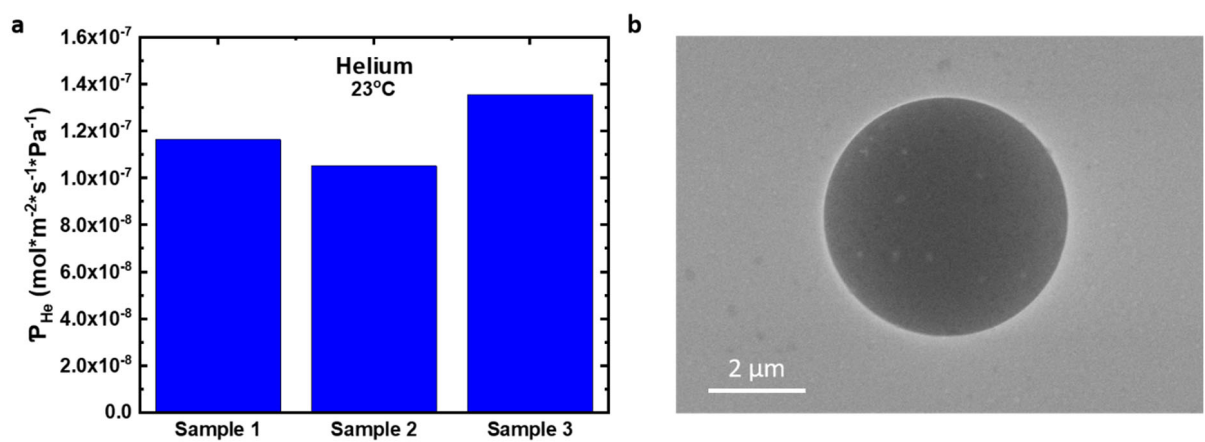

*Figure S3. CNM quality and integrity. a Comparison of permeances of helium between 3 different samples. b Scanning electron microscope image of one of the samples. The CNM looks completely homogeneous without any visible defects.*

### 4. Analysis of the measurements

The software of the mass spectrometer gave the signal measured in millibar units representing the partial pressure of the tested gas in the UHV chamber. This value reflected the equilibrium state between gas permeation into the chamber and constant pumping out of it. It was impossible however to directly calculate gas permeance from the measured signal due to the unknown parameters: influence of the shape of the chamber on the pumping speed, actual pumping speed of the turbomolecular pump for a given gas at given pressure. To overcome this problem, we measured gas flow through a series of open holes with known areas (Figure S4).

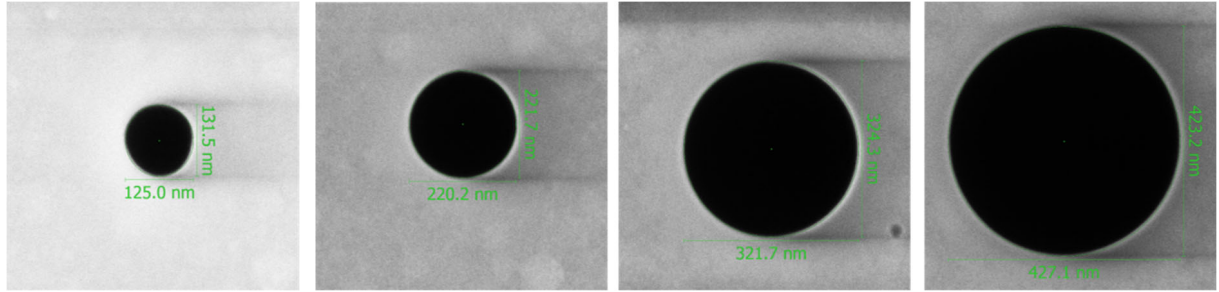

*Figure S4. Calibration samples. These 4 samples were used to calibrate gas permeances. The average diameters are: 127nm, 221nm, 323nm and 425nm respectively.*

The MS signal for these samples could be related to the gas permeance by assuming that the MS response is linear against the gas flow rate. In that case the gas permeance through an open hole could be calculated as the impingement rate of an ideal gas divided by pressure:

$$\mathcal{P}_{ref} = \frac{1}{\sqrt{2\pi mkT} * N_a}$$

Where $\mathcal{P}_{ref}$ is the gas permeance through the calibration sample, m is the molecular mass, k is the Boltzmann constant, T is the absolute temperature and $N_a$ is the Avogadro number. Thus, the normalized MS signal per unit of pressure and area is proportional to the calculated permeance:

$$\frac{I_{ref}}{P_{ref} * A_{ref}} \sim \frac{1}{\sqrt{2\pi mkT} * N_a}$$

Here, I is the MS signal as recorded by the device at a given gas pressure, P is the gas pressure and A is the aperture area. The unknown gas permeance through a CNM sample is also proportional to the normalized MS signal:

$$\frac{I_{sample}}{P_{sample} * A_{sample}} \sim \mathcal{P}_{sample}$$

These relations allow us to calculate permeance for any tested gas using the equation:

$$\mathcal{P}_{sample} = \frac{I_{sample}}{P_{sample} * A_{sample}} \div \frac{I_{ref}}{P_{ref} * A_{ref}} * \frac{1}{\sqrt{2\pi mkT} * N_a} = A \div B * C$$

$I_{sample}/P_{sample}$ and $I_{ref}/P_{ref}$ were calculated by measuring MS signals at different pressures and plotting them against each other (Figure S5).

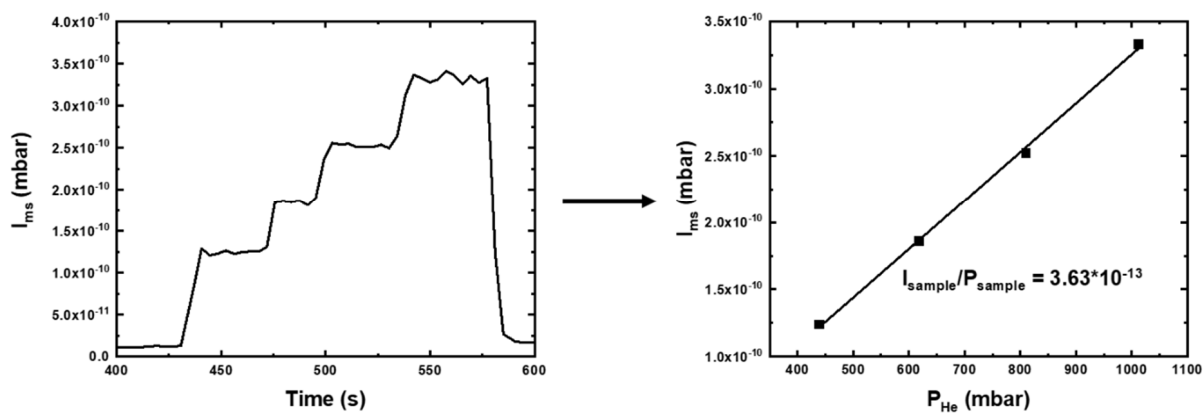

*Figure S5. The analysis of helium permeation.* *The left plot shows the raw data from the measurement of helium permeation through one of the TPT samples at room temperature. "Steps" in the plot correspond to different helium pressures. The right plot shows linear approximation between MS signal and pressure.*

Term A was then calculated by dividing $I_{sample}/P_{sample}$ by aperture's area and term B was calculated by plotting $I_{ref}/P_{ref}$ against calibration aperture areas.

It is important to note that water permeance through CNM was a function of its pressure and could not be expressed by a single value. Thus, it is explicitly written where applicable in the main text at which conditions values for water were calculated. Permeance of water through calibration samples was independent of pressure indicating to a different transport mechanism and allowing to establish a single $I_{ref}/(P_{ref}*A_{ref})$ value.

## 5. Calculations of activation energy of diffusion

Measurements of gas permeances at different pressures allowed us to directly extract apparent activation energies for different gases. This was done by plotting permeance against temperature in Arrhenius coordinates. The obtained values were in fact a sum of activation energy of diffusion and enthalpy of adsorption:

$$E_a = E_{diff} + \Delta H_{ads}$$

$\Delta H_{ads}$ of gases was never measured on CNMs according to our knowledge, so we searched in literature for adsorption of gases on other carbon materials. The values we have found as well as the calculated $E_{diff}$ are presented in the Table S1.

*Table S1. Activation energy of diffusion for different gases.* The table shows energy values for each gas. $E_a$ – the apparent activation energy obtained from experiments, $\Delta H_{ads}$ – the enthalpy of adsorption taken for carbon materials from literature, $E_{diff}$ – activation energy of diffusion was calculated according to the equation in section 5.

|  | $D_{kin}$, pm | $E_a$, kJ/mol | $\Delta H_{ads}$, kJ/mol | $E_{diff}$, kJ/mol |
|---|---|---|---|---|
| He | 260[2] | 26.8 ± 1.2 | -1.5[3] | 28.3 |
| Ne | 275[4] | 26.6 ± 1.3 | -3.7[3] | 30.3 |
| Ar | 340[4] | 40.8 ± 1.8 | -6.6[5] | 47.4 |
| $D_2$ | 289[6] | 27.6 ± 1.8 | -8.3[7] | 35.9 |
| $CO_2$ | 330[6] | 24.0 ± 0.7 | -19[8, 9] | 43.0 |
| $O_2$ | 346[6] | 32.4 ± 1.3 | -17[10] | 49.4 |

The heats of adsorption of helium and neon were measured at 17.3K and 29K respectively on graphitized carbon[3]. We assumed that the obtained mean values of -1.5 and -3.7 kJ/mol depend weakly on temperature and have taken them as is for our calculations. The same considerations were applied for the adsorption of argon and hydrogen. Adsorption of carbon dioxide was more difficult to analyze because the reported literature values ranged between 10 and 29 kJ/mol depending on the type of material. We have taken the $\Delta H_{ads}$ value of 19 kJ/mol because it was characteristic for one of the measured samples and allowed the calculated $E_{diff}$ to fit into the general trend against kinetic diameters. It is however entirely possible that the $\Delta H_{ads}$ of carbon dioxide on TPT CNM is lower or higher and its exact value remains a speculation until a direct measurement is performed. The $\Delta H_{ads}$ of oxygen on charcoal was found to depend on coverage but reached constant value of 17 kJ/mol upon saturation. We have taken this value as a reasonable approximation. The resulting plot $E_{diff}$-$D_{kin}$ did not allow us to make any conclusions about the exact character of this dependence because precise values of $\Delta H_{ads}$ were not known for each gas. However, the obvious presence of the correlation indicated towards the proposed adsorption-diffusion mechanism.